\documentclass[12pt,draftclsnofoot,peerreviewca,onecolumn,a4]{IEEEtran}
\usepackage{amsmath,amssymb,dsfont,stfloats,color,url}
\usepackage[pdftex]{graphicx}
\usepackage{subfigure}
\usepackage{cite}

\DeclareGraphicsExtensions{.eps,.pdf,.png,.jpg,.gif,.jpeg,.pstex}

\newfont{\bbb}{msbm10 scaled 700}

\newfont{\bb}{msbm10 scaled 1100}
\newcommand{\CC}{\mbox{\bb C}}

\newcommand{\ZZ}{\mbox{\bb Z}}

\newcommand{\EE}{\mbox{\bb E}}

\newcommand{\av}{{\bf a}}
\newcommand{\bv}{{\bf b}}

\newcommand{\hv}{{\bf h}}

\newcommand{\sv}{{\bf s}}

\newcommand{\uv}{{\bf u}}
\newcommand{\wv}{{\bf w}}
\newcommand{\vv}{{\bf v}}
\newcommand{\xv}{{\bf x}}
\newcommand{\yv}{{\bf y}}
\newcommand{\zv}{{\bf z}}

\newcommand{\Gm}{{\bf G}}
\newcommand{\Hm}{{\bf H}}
\newcommand{\Id}{{\bf I}}

\newcommand{\Sm}{{\bf S}}

\newcommand{\Wm}{{\bf W}}

\newcommand{\Ym}{{\bf Y}}

\newcommand{\Cc}{{\cal C}}

\newcommand{\Ec}{{\cal E}}

\newcommand{\Ic}{{\cal I}}

\newcommand{\Nc}{{\cal N}}

\newcommand{\Xc}{{\cal X}}

\newcommand{\eqdef}{\stackrel{\Delta}{=}}

\newcommand{\herm}{{\sf H}}

\newcommand{\SINR}{{\sf SINR}}
\newcommand{\SNR}{{\sf SNR}}

\newcommand{\BLUE}{\color[rgb]{0,0,0.90}}

\usepackage{hyperref}
\hypersetup{
    bookmarks=true,         
    unicode=false,          
    pdftoolbar=true,        
    pdfmenubar=true,        
    pdffitwindow=false,     
    pdfstartview={FitH},    
    pdfnewwindow=true,      
    colorlinks=true,       
    linkcolor=red,          
    citecolor=green,        
    filecolor=blue,      
    urlcolor=blue           
}

\usepackage{flushend}
\usepackage{cite}
\usepackage{subfigure}
\usepackage{url}
\usepackage{array}
\usepackage{verbatim}
\usepackage{comment}
\usepackage{stfloats}

\usepackage{color}
\definecolor{red}{rgb}{1,0,0}
\definecolor{blue}{rgb}{0,0,1}

\newtheorem{lemma}{Lemma}

\def\QED{\mbox{\rule[0pt]{1.3ex}{1.3ex}}}

\begin{document}
\title{On the Ergodic Rate Lower Bounds with Applications to Massive MIMO}

\author{Giuseppe Caire$^1$
\thanks{$^1$  Communications and Information Theory Group, 
Technische Universit{\"a}t Berlin, 10587 Berlin, Germany. Email: {\tt  caire@tu-berlin.de}}
}

\maketitle

\begin{abstract}
A well-known lower bound widely used in the massive MIMO literature hinges on {\BLUE {\em channel hardening}}, i.e., 
the phenomenon for which, thanks to the large number of antennas, the {\em effective channel coefficients} resulting from beamforming 
tend to deterministic quantities. If the channel hardening effect does not hold sufficiently well, this bound 
may be quite far from the actual achievable rate. {\BLUE In recent developments of massive MIMO, several scenarios where 
channel hardening is not sufficiently pronounced  have emerged. These settings include, for example, the case of small scattering 
angular spread, yielding highly correlated channel vectors,  and the case of {\em cell-free massive MIMO}. 
In this short contribution, we present two new bounds on the achievable ergodic rate that
offer a complementary behavior with respect to the classical bound: while the former 
performs well in the case of channel hardening and/or when the system is interference-limited 
(notably, in the case of finite number of antennas and {\em conjugate beamforming} transmission), 
the new bounds perform well when the useful signal coefficient does not harden but the channel coherence 
block length is large with respect to the number of users, and in the case where interference is nearly entirely eliminated by 
zero-forcing beamforming. Overall,  using the most appropriate bound depending on the 
system operating conditions yields a better understanding  of the actual performance of systems where channel hardening 
may not occur, even in the presence of a very large number of antennas.}  
\end{abstract}

\begin{IEEEkeywords}
Massive MIMO, achievable ergodic rate, information theoretic bounds.
\end{IEEEkeywords}

\newpage

\section{Introduction}  \label{intro}

Multiuser MIMO (MU-MIMO) and its large-antenna regime embodiment known as {\em massive MIMO} \cite{Larsson-book}  is one of the most promising 
technologies to achieve very high spectral efficiency in wireless networks. 
Massive MIMO  has been very intensively studied in the past few years and it is still a very active research topic (e.g., see \cite{Larsson-book} and references therein).
Furthermore, massive MIMO based on TDD reciprocity for the estimation of the downlink (DL) channel 
vectors from uplink pilot signals  has been demonstrated in practice in several academic and industrial  
prototypes \cite{argos,larsson2014massive,forenza2015achieving},  thus {\BLUE confirming the possibility of obtaining accurate 
and timely DL channel estimates at the base station side 
from pilot symbols sent by the users in the uplink direction.} 

Restricting to linear beamforming (for DL transmission), {\BLUE single data stream per user}, and independent channel coding 
of the user data streams,  a generic channel use of the underlying channel model is described by the Gaussian interference channel:
\begin{equation} \label{interfch}
y_k = g_{k,k} s_k + \sum_{k' \neq k} g_{k,k'} s_{k'} + z_k, \;\;\;\; k = 1, \ldots, K
\end{equation}
where $y_k$ is the channel output observed by user $k$ decoder, 
$s_k$ is the coded information bearing symbol for user $k$ (useful signal), $z_k$ is AWGN, and $\{g_{k,k'}\}$ are the 
{\em effective channel coefficients} resulting from {\BLUE the inner products of the transmit beamforming vectors with 
the users' channel vectors.} 

{\BLUE In general, when the coefficients $\{g_{k,k'}\}$ are not known to user $k$ receiver, 
it is not clear what is ``signal'' and what is ``interference''  in the signal model (\ref{interfch}). In particular, the intuitive notion of 
Signal-to-Interference plus Noise Ratio, given by  $\SINR_k = \frac{|g_{k,k}|^2}{N_0 + \sum_{k'\neq k} |g_{k,k'}|^2}$, is in general 
not rigorously related to a corresponding notion of information theoretic achievable rate. 

When coding is performed across many channel states, under standard assumptions 
on the joint stationarity and ergodicity of the channel coefficients and the CSI, the  
relevant notion of achievable rate is usually referred to as {\em ergodic achievable rate} \cite{BPS}.} 
In \cite{Larsson-book}  (see also the many references therein), a rigorous lower bound is derived
for the ergodic achievable rate of MU-MIMO systems with effective channel (after DL beamforming) represented by (\ref{interfch}).  
This lower bound works well when the useful signal coefficient $g_{k,k}$ behaves almost deterministically, i.e., it has a non-zero mean and a small variance. 
Otherwise,  when Var$(g_{k,k})$ is not negligible with respect to $|\EE[g_{k,k}]|^2$, the bound displays a self-interference limited behavior, i.e., 
as the Signal-to-Noise Ratio (SNR) increases, the bound converges to a finite asymptotic limit instead of growing linearly with  $\log \SNR$. 
Such self-interference behavior has prompted some authors to suggest that knowledge of the useful signal coefficient at the receiver is critically important, and
the use of {\em dedicated beamformed pilots symbols} in the DL transmission in order to enable the estimation of the 
effective channel coefficients at the user receivers has been studied in various works (e.g., see \cite{ngo2013massive,zuo2016multicell,kim2015study}). 

In this short paper, we derive {\BLUE two} new lower bounds on the achievable ergodic rate of MU-MIMO systems, or more in general, 
Gaussian interference channels with unknown channel coefficients at the receiver, with suitable side information, and under the constraint 
of treating interference as noise \cite{TIN}. {\BLUE In addition, for the sake of being self-contained, we also derive 
a max-min upper bound (Lemma \ref{UB-lemma}) and present the derivation of the widely used in the massive MIMO literature and described in \cite{Larsson-book}
(Lemma \ref{LB1-lemma}). This will be useful to make comparisons with the two new lower bounds in this paper. 
We shall show through examples that the two new lower bounds (Lemma \ref{LB2-lemma} and Lemma \ref{LB3-lemma}) 
have a  complementary behavior with respect to the commonly used bound (Lemma \ref{LB1-lemma}). In particular, they are 
able to closely follow the upper bound (Lemma \ref{UB-lemma}) even without significant channel hardening, 
provided that channel coherence block length (defined in Section \ref{model-notation}) is large with respect to the number of users and the DL 
beamforming is able to significantly remove the multiuser interference (e.g., in the presence of zero-forcing beamforming). 
Therefore, our bounds somehow corroborate the fact that beamformed DL pilots are indeed not critically needed even in the cases where 
the useful signal coefficients suffers from significant statistical fluctuations, despite the large number of antennas. 

In recent developments of massive MIMO, several scenarios where 
channel hardening is not sufficiently pronounced  have emerged. These settings include, for example, the case of sparse support of the 
channel angular scattering function, yielding highly correlated channel vectors,  and the case of {\em cell-free massive MIMO} \cite{ngo2015cell}, where
antennas are spatially distributed over a large area and only a relatively small number of antennas have significant large-scale 
channel strength with respect to any given user. 
The case of highly correlated channel vectors received a lot of attention 
motivated by propagation models for mm-waves and by the opportunity of exploiting the channel sparsity 
in order to use {\em compressed sensing} techniques for channel estimation with reduced pilot overhead 
(e.g., see \cite{mahdi-FDD,adhikary2013joint,adhikary2014joint,nam2017role,rao2014distributed,fang2017low}).
Also, during the revision of this paper, the bounding technique in Lemma \ref{LB2-lemma}, taken from our ArXiv preprint 
\cite{caire2017ergodic}, was used in \cite{chen2017channel} to provide a more accurate performance analysis of cell-free massive 
MIMO.  In both the cell-free and the highly correlated channel cases, channel hardening is not so pronounced and 
the new bounds in this paper may provide a useful alternative tool for accurate system performance evaluation.}

{\BLUE In order to put these bounds in a historical perspective, we observe that  the bound in  \cite{Larsson-book}  
(Lemma \ref{LB1-lemma}) has  been ``re-discovered'' many times in different contexts. 
To the best of the author's knowledge, this bounding technique appeared first 
in the work of Medard \cite[Eq. 40-46]{medard2000effect}. 
The bounding approach used here to derive the new bounds in Lemma \ref{LB2-lemma} and \ref{LB3-lemma} can be traced back to
the tutorial paper by Biglieri, Proakis, and Shamai \cite{BPS}, in particular Eq. 3.3.27 and 3.3.60. Both our Lemmas follow by 
neglecting one term in the expansion of the mutual information found in these equations, and further manipulating 
the remaining terms in order to obtain an easily computable bound.} 


{\BLUE The fact that the self-interference of the classical bound in Lemma \ref{LB1-lemma} is more an artifact of the bounding technique than 
a fundamental system limit has some interesting consequences from the design of massive MIMO systems. 
In particular, DL transmission resources should not be wasted by sending beamformed  pilot symbols. 
Attractive alternatives consists of using blind estimation schemes (e.g., as in \cite{Larsson-blind}) or codes for the non-coherent block-fading
channel (e.g., \cite{marzetta1999capacity,divsalar1990multiple}). Furthermore, our results point out that, as long as we can afford 
coding across sufficiently many time and frequency blocks, such that the ergodic regime is relevant, 
channel hardening is not very important and massive MIMO can be used very successfully even in cases of 
sparse scattering.}

\section{Notation and model assumptions}  \label{model-notation}

Consider a basic MU-MIMO system with $M$ antennas at the base station and $K$ single-antenna users.\footnote{{\BLUE Often 
the model in (\ref{ziocane}) is referred to as ``MISO'' system (meaning Multiple-Input Single-Output). 
We prefer to refer to it as {\em MU-MIMO with single-antenna users} since the system has indeed multiple inputs 
($M$ base station antennas) and multiple outputs ($K$ users antennas), although the outputs are not jointly processed.}}
A channel use of the DL channel can be represented as
\begin{equation} \label{ziocane}
y_k = \hv_k^\herm \xv + z_k, \;\;\; k = 1, \ldots, K
\end{equation}
where $\Hm = [\hv_1, \ldots, \hv_K]$ is $M \times K$ channel matrix whose columns 
represent the propagation channels between the $M$ base station antennas to each user antenna, and $z_k$ is the sample of an AWGN process
with components $\sim \Cc\Nc(0,N_0)$. 
This channel model is relevant for an OFDM system where (\ref{ziocane}) describes a single time-frequency symbol. 
In general,  {\BLUE the} channel bandwidth of $W$ Hz is divided into {\em coherence subbands} of width $W_s$, over which the channel coefficients 
are {\BLUE frequency-invariant}, and the time axis is divided into {\em coherence blocks} of duration $T_s$, over which the channel coefficients 
are time-invariant. A coherence block in the time-frequency plane {\BLUE consists of} a tile of size $W_s \times T_s$
over which the channel coefficients are essentially both frequency and time invariant. 
The number of channel uses {\BLUE (i.e., signal space dimensions in the time-frequency domain)} 
spanning a coherence block is $T \approx  \lceil W_s T_s \rceil$. 
This approximation {\BLUE is usually referred to as ``block-fading model''} 
and it is widely used in the wireless communications and information theoretic literature. In addition, 
it is also a very good approximation for all practical purposes for wireless systems based on OFDM. In fact, if the block-fading model
does not approximately hold, an OFDM system would be affected by severe inter-carrier interference and the simple discrete parallel channel 
model for OFDM would not apply any longer.  Consistently with an exceedingly large number of works in this area, we shall assume that the block fading model
holds exactly.  We do not assume that the channel coefficients are independent across different coherence blocks. In fact, they can be strongly correlated.
However, we assume that the channel matrix process $\{\Hm[t] : t  \in \ZZ\}$, where $t$ counts the coherence blocks, is a stationary ergodic process. 
In order to indicate the fact that we have $T$ channel uses per coherence block, we shall write 
\begin{equation} \label{ziogatto}
y_k[tT:(t+1)T-1] = \hv^\herm_k[t] \xv[tT:(t+1)T-1] + z_k[tT:(t+1)T-1], \;\;\; k = 1, \ldots, K
\end{equation}
where $y_k[tT:(t+1)T-1] \in \CC^{1 \times T}$, $\xv[tT:(t+1)T-1] \in \CC^{M \times T}$, and $z_k[tT:(t+1)T-1] \in \CC^{1 \times T}$ 
are the received, transmitted, and noise {\BLUE {\em supersymbols}, i.e., signal blocks formed by $T$ channel uses each}, 
and $\hv_k[t]$ is the $k$-th column of the channel matrix $\Hm[t]$ on coherence block $t$. 


We consider linear precoded transmission to the $K$ users, where $K$ independent messages are sent to the $K$ users over multiple coherence blocks. 
The $K$ codewords are divided in blocks of $T$ symbols, and on each coherence block $t$ they are jointly precoded 
and transmitted over the MU-MIMO channel.  The transmitted signal supersymbol on coherence block $t$ is given by 
\begin{equation} \label{xx}
\xv[tT:(t+1)T-1]  = \sum_{k=1}^K \sqrt{\Ec_k[t]} \vv_k[t] s_k[tT:(t+1)T-1], 
\end{equation}
where $\vv_k[t] \in \CC^{M \times 1}$ is the precoding vector for user $k$ in block $t$,  
$s_k[tT:(t+1)T-1] \in \CC^{1 \times T}$ is the corresponding coded data-bearing signal block of user $k$. 
We assume unit precoding vectors, i.e., 
$\|\vv_k[t]\|^2 = 1$ for all $k$ and $t$, and we assume that the user codewords satisfy 
$\frac{1}{T} \EE[ \|s_k[tT:(t+1)T-1]\|^2] = 1$, where the expectation is taken over the codebook, 
with uniform probability over the codewords.  With these normalizations, $\Ec_k[t]$ represents the transmitted energy per channel use 
for the data stream of user $k$ on supersymbol $t$. 

For the sake of the following analysis, it does not really matter how the vectors $\{\vv_k[t]\}$ and the transmitted energies per symbol $\{\Ec_k[t]\}$ are determined, 
as long as they are independent of the codewords $\{s_k[tT:(t+1)T-1]\}$ and of the additive noise process $z_k[tT:(t+1)T-1]$. 
This assumption is normally always verified  since the codewords are determined by fixing a codebook for each user, and selecting the individual information messages independently 
of anything else,  with uniform probability. It is also obvious that the precoding vectors and allocated transmit energy per symbol on block $t$ is independent of the noise realization 
on block $t$, which is unknown at the transmitter side. 
However, we allow  $\{\vv_k[t]\}$ and $\{\Ec_k[t]\}$ to be functions of the channel matrix process $\{\Hm[t]\}$ and possibly of other correlated 
processes, e.g., arising from some form of channel measurement, causal feedback, quantization, or TDD reciprocity mechanism, as long as they are determined 
at the beginning of the block and kept fixed over each block, and as long as the processes $\{\Hm[t], \vv_1[t], \ldots, \vv_K[t], \Ec_1[t], \ldots, \Ec_K[t]\}$ 
are jointly stationary and ergodic.

\section{Achievable rate bounds}  \label{rate-bounds}

In this section we present a simple upper bound and {\BLUE 
three simple lower bounds} to the achievable ergodic rate for user $k$
in the previously defined MU-MIMO DL system. 
{\BLUE As anticipated in Section \ref{intro}, the upper bound (Lemma \ref{UB-lemma}) 
and the first lower bound (Lemma \ref{LB1-lemma}) are well-known. They are presented here for the sake of completeness, 
and since it may be useful to have them all in a single place, developed in a consistent notation.
The other lower bounds (Lemma \ref{LB2-lemma} and \ref{LB3-lemma}) are somehow new, 
or at least not well-known in the massive MIMO literature, as discussed in more details in Section \ref{intro}.}

Replacing (\ref{xx}) into (\ref{ziogatto}) we can write the received signal block at user $k$ decoder as
\begin{equation}  \label{ziominchia}
y_k[tT:(t+1)T-1] = \sum_{k'=1}^K \sqrt{\Ec_{k'}[t]} \left ( \hv^\herm_{k}[t] \vv_{k'}[t]\right ) s_{k'}[tT:(t+1)T-1] + z_k[tT:(t+1)T-1] .
\end{equation}
Standard information theory results yield that user $k$ can achieve rate
\begin{equation} \label{minfo}
R_k = \frac{1}{T} I \left ( s_k[1:T] ; y_k[1:T] \right ),  
\end{equation}
{\BLUE where $s_k[1:T]$ and $y_k[1:T]$ have the joint marginal statistics of the corresponding 
$t$-th supersymbols in (\ref{ziominchia}) and $R_k$ is rate is expressed in bit per channel use, due to the normalization of the
block-wise mutual information by the number of channel uses per block $T$.} 
The block-wise model (\ref{ziogatto}) is not generally memoryless since there may be memory between the blocks due to the fact that 
$\{\Hm[t], \vv_1[t], \ldots, \vv_K[t], \Ec_1[t], \ldots, \Ec_K[t]\}$ may be correlated over time (i.e., over the sequence of blocks). 
However, cutting the channel into blocks, treating the blocks as supersymbols, and neglecting the memory between them, 
yields a possibly suboptimal achievable rate.  It is clear that in the mutual information (\ref{minfo}) only the first-order marginal distribution of the 
processes $\{\Hm[t], \vv_1[t], \ldots, \vv_K[t], \Ec_1[t], \ldots, \Ec_K[t]\}$ plays a role. 
An important observation here is that the mutual information expression in (\ref{minfo}) implicitly implies that that the decoder of user $k$ treats the multiuser 
interference as additional additive noise, i.e., the rate in (\ref{minfo}) is achieved by {\em Treating Interference as Noise} (TIN) \cite{TIN}. 
{\BLUE Of course, this noise may be treated as non-Gaussian 
(e.g., see \cite{dytso2014gaussian}), 
by incorporating in the decoder the available {\em a priori} information that user $k$ receiver has about the interference caused by the signals of users $k' \neq k$.} 
Finally, it is also implicit in (\ref{minfo}) that there is no assumed or genie-aided CSI at the user $k$ decoder, in fact the mutual information in (\ref{minfo}) has no 
conditioning with respect to any additional  ``channel state information''  variable. 

We start with an upper bound in the max-min sense:

\begin{lemma} \label{UB-lemma}
Under the system assumptions defined before, {\BLUE the max-min of $R_k$ in (\ref{minfo}),} where the max is over the coding/decoding strategy 
of user $k$ and the min is over all input distributions of the other users $k' \neq k$, is upper-bounded by
\begin{equation}  \label{UB}
R^{\rm ub}_k = \EE \left [  \log \left ( 1 + \frac{|g_{k,k}|^2}
{ N_0 + \sum_{k' \neq k} |g_{k,k'}|^2}    \right ) \right ],
\end{equation}
where the random variables $\{g_{k,k'} : k' = 1,\ldots, K\}$ have the same joint first-order marginal distribution of 
\[ \left \{ \sqrt{\Ec_{k'}[t]} \left  ( \hv^\herm_k[t] \vv_{k'}[t]\right ) : k' = 1, \ldots, K \right \}. \]
\hfill $\square$
\end{lemma}

{\em Proof.}  
{\BLUE Omitting the block index $t$, a single supersymbol of the model in (\ref{ziominchia}) can be written concisely as}
\begin{equation}  \label{slot}
y_k[1:T] = g_{k,k} s_k[1:T] + \sum_{k' \neq k} g_{k,k'} s_{k'}[1:T] + z_k[1:T],
\end{equation}
where we define
\begin{equation} \label{gcoef}
g_{k,k'} =  \sqrt{\Ec_{k'}}    \left ( \hv^\herm_{k}  \vv_{k'}\right ).
\end{equation}
{\BLUE For three random variables $X,Y,Z$ with joint probability distribution $P_{X,Y,Z} = P_X P_Z P_{Y|X,Z}$ is it immediate to show
that
\begin{equation} \label{xyz}
I(X;Y) \leq I(X;Y|Z). 
\end{equation}
Let's indicate $\{g_{k,k'}:k' = 1,\ldots, K\}$ briefly as $\{g_{k,k'}\}$. 
Using the fact that $\{g_{k,k'}\}$ and the input $s_{k}[1:T]$ are statistically independent, using (\ref{xyz}) we can write}
\[ I \left ( s_k[1:T] ; y_k[1:T] \right )   \leq I \left ( s_k[1:T] ; y_k[1:T] | \{g_{k,k'}\} \right ) \]
For given $\{g_{k,k'}\}$, the worst-case additive interference subject to a power constraint in (\ref{slot}) 
is obtained by letting $s_{k'}[1:T]$ to be i.i.d. with components $\sim \Cc\Nc(0,1)$ for all $k' \neq k$ \cite{hassibi2003much}. 
At this point, we are in the presence of a Gaussian additive noise channel (conditionally on $\{g_{k,k'}\}$) with channel state and noise variance 
known at the receiver, and varying over blocks of length $T$ symbols according to a stationary ergodic process. 
The capacity in bits per block of such channel is immediately given by 
\begin{equation} 
T \EE \left [ \log \left ( 1 + \frac{|g_{k,k}|^2}{N_0 + \sum_{k' \neq k} |g_{k,k'}|^2} \right ) \right ]
\end{equation}
Dividing by $T$ we obtain (\ref{UB}).  
\hfill \QED

It is interesting to remark that expression (\ref{UB}) has been often referred to as the achievable ergodic rate, in the presence of
perfect knowledge of the channel coefficients $\{g_{k,k'}\}$ at receiver $k$. This is indeed correct, but this is not in general the best achievable rate
even insisting on linear precoding and TIN. In fact, fixing the linear precoding scheme, we are in the presence of a Gaussian $K \times K$ interference channel
with coefficients $\{g_{k,k'}\}$, for which the Gaussian input distribution is generally not optimal, even under TIN \cite{dytso2014gaussian}. Under certain conditions of
weak interference, Gaussian inputs are indeed approximately optimal as shown in \cite{TIN}. In practice, when the MU-MIMO linear precoding is effective, 
the crosstalk coefficients $g_{k,k'}$ for $k' \neq k$ are much weaker than the useful signal coefficients $g_{k,k}$ and 
the use of Gaussian inputs is fully justified. 

The following lower bound is widely used in the massive MIMO literature (e.g., see \cite{Larsson-book}).

\begin{lemma} \label{LB1-lemma}
The ergodic achievable rate $R_k$ in (\ref{minfo}) is lower bounded by
\begin{equation}  \label{LB1}
R^{\rm lb1}_k = \log \left ( 1 + \frac{\left | \EE\left [ g_{k,k} \right ] \right |^2 }
{N_0 +  {\rm Var} \left ( g_{k,k} \right )
+ \sum_{k' \neq k} \EE \left [ |g_{k,k'}|^2 \right ]}  \right ). 
\end{equation}
where the random variables $\{g_{k,k'} : k' = 1,\ldots, K\}$ are defined as in Lemma \ref{UB-lemma}.
\hfill $\square$
\end{lemma}

{\em Proof.} 
Consider again the supersymbol channel model in (\ref{slot}). 
Since we are after a lower bound, we can choose a suitable input distribution to lower bound the mutual information. 
In particular, here we set all user inputs to be Gaussian with i.i.d. components $\sim \Cc\Nc(0,1)$. Then, we can write
\begin{eqnarray} 
I \left ( s_k[1:T] ; y_k[1:T] \right )  & = & h(s_k[1:T]) - 
h\left ( s_k[1:T] | y_k[1:T] \right ) \nonumber \\
& = &  h(s_k[1:T]) -  h\left (s_k[1:T] - \widehat{s}_k[1:T] | y_k[1:T]\right ) \label{mmse} \\
& \geq & T \left ( \log (\pi e) - \log (\pi e {\sf MMSE}(S|Y)) \right ) \label{mmse2}
\end{eqnarray}
where in (\ref{mmse}) we used the fact that differential entropy is invariant to constant shifts of the probability density, 
and we let $\widehat{s}_k[1:T]$ to be the linear symbol-by-symbol MMSE estimator of the sequence
$s_k[1:T]$ from the observation $y_k[1:T]$, which is therefore a function of  $y_k[1:T]$, and where 
(\ref{mmse2}) follows from the fact  removing conditioning does not reduce the differential entropy, and that the complex circularly symmetric 
Gaussian distribution is a differential entropy maximizer for given second moment, where the quantity 
${\sf MMSE}(S|Y)$ indicates the per-component Mean-Square Error of the linear symbol-by-symbol MMSE estimator of
$s_k[1:T]$ from $y_k[1:T]$. Let $s_k$ and $y_k$ denote generic components of $s_k[1:T]$ and $y_k[1:T]$, respectively. Standard calculations yield the estimator
\[ \widehat{s}_k(y_k) = \frac{\EE[s_k y^*]}{\EE[|y_k|^2]} y_k \]
yielding the MSE
\begin{eqnarray} 
{\sf MMSE}(S|Y) & = & \EE[|s_k|^2] - \frac{|\EE[ s_k y^*_k] |^2}{\EE[|y_k|^2]} \nonumber \\
& = & 1 - \frac{|\EE[g_{k,k}]|^2}{N_0 + \EE[|g_{k,k}|^2] +  \sum_{k' \neq k} \EE[|g_{k,k'}|^2]} \nonumber \\
& = & \frac{N_0 + {\rm Var}(g_{k,k}) +  \sum_{k' \neq k} \EE[|g_{k,k'}|^2]}{N_0 + \EE[|g_{k,k}|^2] +  \sum_{k' \neq k} \EE[|g_{k,k'}|^2]}  \label{mmse-trick}
\end{eqnarray}
noticing that $\EE[|g_{k,k}|^2] = {\rm Var}(g_{k,k}) + |\EE[g_{k,k}]|^2$. Replacing (\ref{mmse-trick}) into 
(\ref{mmse2}), dividing by $T$ we arrive at (\ref{LB1}). 
\hfill \QED

{\BLUE Next, we present our first new lower bound.}

\begin{lemma} \label{LB2-lemma}
The ergodic achievable rate $R_k$ in (\ref{minfo}) is lower bounded by
\begin{equation}  \label{LB2}
R^{\rm lb2}_k = 
\EE \left [  \log \left ( 1 + \frac{|g_{k,k}|^2}{N_0 + \sum_{k' \neq k} |g_{k,k'}|^2}    \right ) \right ]  
- \frac{1}{T} \sum_{k'=1}^K \log \left (1 + \frac{T}{N_0} {\rm Var}(g_{k,k'}) \right ),
\end{equation}
where the random variables $\{g_{k,k'} : k' = 1,\ldots, K\}$ are defined as in Lemma \ref{UB-lemma}. 
\hfill $\square$
\end{lemma}

{\em Proof.}  
We start again from the supersymbol channel (\ref{slot}). Choosing Gaussian independent input distributions for the codewords
and using the chain rule of mutual information, we can write
\begin{eqnarray}
I \left ( \{g_{k,k'}\} , s_k[1:T] ; y_k[1:T] \right ) & = & I \left (s_k[1:T] ; y_k[1:T] \right ) + I \left (\{g_{k,k'}\} ; y_k[1:T] | s_k[1:T]  \right ) \nonumber \\
& = & I \left ( \{g_{k,k'}\}  ; y_k[1:T] \right )  + I \left ( s_k[1:T] ; y_k[1:T] | \{g_{k,k'}\} \right ),  
\end{eqnarray}
from which we can write
\begin{eqnarray} 
I \left ( s_k[1:T] ; y_k[1:T] \right ) & = &
I \left ( s_k[1:T] ; y_k[1:T] | \{g_{k,k'}\} \right ) \nonumber \\
& & - I \left (\{g_{k,k'}\}  ; y_k[1:T] | s_k[1:T]  \right ) +  I \left (\{g_{k,k'}\}  ; y_k[1:T]  \right ) \nonumber \\
& \geq & 
I \left ( s_k[1:T] ; y_k[1:T] | \{g_{k,k'}\} \right ) - I \left (\{g_{k,k'}\}  ; y_k[1:T] | s_k[1:T]  \right ).  \label{term0}
\end{eqnarray}
The mutual information $I \left ( s_k[1:T] ; y_k[1:T] | \{g_{k,k'}\} \right )$ is easily lower-bounded by using the worst-case 
additive (uncorrelated) noise result \cite{hassibi2003much}, and yields
\begin{equation} \label{term1}
I \left ( s_k[1:T] ; y_k[1:T] | \{g_{k,k'}\} \right ) \geq 
T \EE \left [ \log\left ( 1 + \frac{|g_{k,k}|^2}{N_0 + \sum_{k' \neq k} |g_{k,k'}|^2  } \right ) \right ].
\end{equation}
{\BLUE Since $\{g_{k,k'}\}$ and $\{s_{k'}[1:T] : k' \neq k \}$ are independent, using (\ref{xyz})
we can upper bound second mutual information term in (\ref{term0}) as}
\begin{align}
& I \left (\{g_{k,k'}\}  ; y_k[1:T] | s_k[1:T]  \right ) \nonumber \\
& \leq   I \left (\{g_{k,k'}\}  ; y_k[1:T] | s_k[1:T] , \{s_{k'}[1:T] : k' \neq k\} \right ) \nonumber \\
& = I \left (\{g_{k,k'}\}  ;  \sum_{k'=1}^K g_{k,k'} s_{k'}[1:T] + z_k[1:T] | \{s_{k'}[1:T] : k' =1,\ldots, K\}  \right ).
 \label{porcodio}
\end{align}
Now, we notice that the mutual information in (\ref{porcodio}) corresponds to a MIMO channel
with {\BLUE $K$-dimensional input} $\{g_{k,k'}\}$, {\BLUE $T$-dimensional output} $\sum_{k'=1}^K g_{k,k'} s_{k'}[1:T] + z_k[1:T]$, and known channel 
matrix $\Sm$ of dimensions $T \times K$, with Gaussian i.i.d. columns given by the vectors $\{s_{k'}[1:T] : k' =1,\ldots, K\}$. 
Using standard results on differential entropy maximization \cite{cover2012elements}, we have that 
the mutual information in (\ref{porcodio}) is maximized by letting $\{g_{k,k'}\}$ jointly Gaussian with the assigned covariance matrix 
$\Gm_k$ with $(\ell,m)$ elements\footnote{{\BLUE Notice that $\Gm_k$ depends on the joint statistics of $\{g_{k,k'}\}$ which is assumed to be known according to 
the model assumptions made at the beginning of Section \ref{model-notation}.}} 
\[ [\Gm_k]_{\ell,m} = \EE[ g_{k,\ell} g^*_{k,m} ] - \EE[g_{k,\ell}] \EE[g^*_{k,m}]. \]
The resulting upper bound is
\begin{equation} \label{porcamadonna}
 I \left (\{g_{k,k'}\}  ; y_k[1:T] | s_k[1:T]  \right )  \leq \EE \left [ \log \left | \Id_K + \frac{1}{N_0} \Sm^\herm \Sm \Gm_k \right | \right ]. 
\end{equation}
This upper bound is enough to get a rate lower bound by using 
(\ref{term1}) and (\ref{porcamadonna}) in (\ref{term0}). However, it requires the computation of the $K \times K$ matrix $\Gm_k$ 
and the expectation of the log-det formula with respect to the central Wishart matrix $\Sm^\herm \Sm$. In order to obtain a simpler (but looser) 
bound, we can use Jensen's inequality to the concave log-det function and notice that 
$\EE[\Sm^\herm \Sm] = T \Id_K$. This yields
\begin{equation} \label{porcamadonna1}
 I \left (\{g_{k,k'}\}  ; y_k[1:T] | s_k[1:T]  \right )  \leq \log \left | \Id_K + \frac{T}{N_0} \Gm_k \right |. 
\end{equation}
Finally, using Hadamard inequality, we obtain the laxer but simpler bound
\begin{equation} \label{porcamadonna2}
 I \left (\{g_{k,k'}\}  ; y_k[1:T] | s_k[1:T]  \right )  \leq \sum_{k' = 1}^K \log \left (1 + \frac{T}{N_0} {\rm Var}(g_{k,k'})  \right ). 
\end{equation}
Using  (\ref{term1}) and (\ref{porcamadonna2}) in (\ref{term0}) yields (\ref{LB2}). 
\hfill \QED

{\BLUE 
Combining Lemma \ref{UB-lemma} and Lemma \ref{LB2-lemma} we have that in the limit of very large coherence block length $T$ the ergodic rate upper bound
in (\ref{UB} is achievable. This indicates that, irrespective of whether the effective channel coefficients $\{g_{k,k'}\}$ harden to deterministic limits of remain 
random, if they remains constant over time and frequency for a very large number of symbols there is no price to pay for not knowing these coefficients 
at the receiver. Notice that this conclusion cannot be obtained from Lemma \ref{LB1-lemma}, because of the self-interference term 
Var$(g_{k,k})$ at the denominator, which does not depend on the coherence block length $T$. 
As we shall see in the numerical examples of Section \ref{results}, in some cases the 
bound (\ref{LB2}) can be significantly tighter than bound (\ref{LB1}).
In particular, this happens when $T$ is significantly larger than $K$, 
the useful signal coefficient $g_{k,k}$ presents significant statistical fluctuations (lack of hardening), and 
the MU-MIMO beamforming is able to nearly eliminate the multiuser interference (e.g., in the case of Zero-Forcing Beamforming (ZFBF)), such that
the coefficient variances Var$(g_{k,k'})$ for $k' \neq k$ are small.  However, the bound (\ref{LB2}) has an annoying drawback: 
when Var$(g_{k,k'})$ are fixed quantities, independent of SNR, and $T$ is fixed, 
the bound (\ref{LB2}) becomes completely useless and, in fact, can take on negative values for sufficiently 
large SNR. 

Next, we present another lower bound that does not suffer from this problem, although it is slightly 
more complicated for numerical evaluation.

\begin{lemma} \label{LB3-lemma}
The ergodic achievable rate $R_k$ in (\ref{minfo}) is lower bounded by
\begin{align}  \label{LB3}
R^{\rm lb3}_k & = 
\EE \left [  \log \left ( 1 + \frac{|g_{k,k}|^2}{N_0 + \sum_{k' \neq k} \EE[ |g_{k,k'}|^2 | g_{k,k}] }    \right ) \right ]  
- \frac{1}{T} \log \left (1 + \frac{T {\rm Var}(g_{k,k})}{N_0 + \sum_{k' \neq k} \EE[|g_{k,k'}|^2]}  \right ) \nonumber \\
& + \EE\left [ \log \left ( 1 + \frac{1}{N_0} \sum_{k'\neq k} |g_{k,k'}|^2 \right ) \right ]  -  \log \left ( 1 + \frac{1}{N_0} \sum_{k'\neq k} \EE[|g_{k,k'}|^2] \right ) 
\end{align}
where the random variables $\{g_{k,k'} : k' = 1,\ldots, K\}$ are defined as in Lemma \ref{UB-lemma}. 
\hfill $\square$
\end{lemma}


{\em Proof.}  
We start again from the supersymbol channel (\ref{slot}). Choosing Gaussian independent input distributions for the codewords
and proceeding as in the beginning of the proof of Lemma \ref{LB2-lemma},  we can write
\begin{eqnarray}
I \left ( g_{k,k} , s_k[1:T] ; y_k[1:T] \right ) & = & I \left (s_k[1:T] ; y_k[1:T] \right ) + I \left (g_{k,k} ; y_k[1:T] | s_k[1:T]  \right ) \nonumber \\
& = & I \left ( g_{k,k}  ; y_k[1:T] \right )  + I \left ( s_k[1:T] ; y_k[1:T] | g_{k,k} \right ),  
\end{eqnarray}
from which we have
\begin{eqnarray} 
I \left ( s_k[1:T] ; y_k[1:T] \right ) & \geq &
I \left ( s_k[1:T] ; y_k[1:T] | g_{k,k} \right ) - I \left ( g_{k,k}  ; y_k[1:T] | s_k[1:T]  \right )  \label{term03}
\end{eqnarray}
For given $g_{k,k}$, the channel in (\ref{slot}) is an additive non-Gaussian noise channel with Gaussian input and 
(conditional) uncorrelated noise with (conditional) per-component 
variance given by $N_0 + \sum_{k'\neq k} \EE[|g_{k,k'}|^2 | g_{k,k}]$. Hence, 
applying the worst-case additive (uncorrelated) noise result \cite{hassibi2003much} {\em conditionally} on $g_{k,k}$,
we obtain the lower bound
\begin{equation} \label{term13}
I \left ( s_k[1:T] ; y_k[1:T] | g_{k,k} \right ) \geq 
T \EE \left [ \log\left ( 1 + \frac{|g_{k,k}|^2}{N_0 + \sum_{k' \neq k} \EE[|g_{k,k'}|^2|g_{k,k}]  } \right ) \right ].
\end{equation}
In passing, we notice that we cannot further lower bound this term by using Jensen's inequality and taking the outer expectation with respect to 
$g_{k,k}$ in the denominator inside the log in (\ref{term13}) (thus removing the conditioning in the terms $\EE[|g_{k,k'}|^2|g_{k,k}]$)
since $g_{k,k}$ appears also in the numerator of this term. However, if the case the coefficients $g_{k,k'} : k' \neq k$ are independent of 
$g_{k,k}$, the conditioning disappears and (\ref{term13}) is further simplified.
 
In order to upper bound the second mutual information in (\ref{term03}), we write as follows
\begin{eqnarray}
I \left (g_{k,k}  ; y_k[1:T] | s_k[1:T]  \right ) & = &  h(y_k[1:T] |s_k[1:T]) - h(y_k[1:T] | g_{k,k}, s_k[1:T])  \label{porcodio3}
\end{eqnarray}
We consider each differential entropy in the RHS of (\ref{porcodio3}) separately. 
The first differential entropy can be upper bounded by assuming $y_k[1:T]$ to be conditionally Gaussian given $s_k[1:T]$ with the same (conditional) 
covariance matrix \cite{cover2012elements}. 
For the sake of notation simplicity, let $\yv_k$ denote the $T \times 1$ column vector corresponding to the supersymbol $y_k[1:T]$, and likewise
$\{\sv_k' : k' = 1, \ldots, K\}$ and $\zv_k$ have the same meaning with respect to $\{s_{k'}[1:T] : k' = 1, \ldots, K\}$ and 
$z_k[1:T]$, respectively. 
Then, the conditional covariance of $\yv_k$ given $\sv_k$ is given by 
\begin{eqnarray}
{\rm Cov}(\yv_k | \sv_k) & = & \EE \left [ \yv_k \yv_k^\herm | \sv_k \right ] - \EE \left [ \yv_k |\sv_k \right ] \EE \left [ \yv^\herm_k |\sv_k \right ]  \nonumber \\
& = & \sv_k \sv_k^\herm {\rm Var}(g_{k,k}) + \left ( N_0 + \sum_{k'\neq k} \EE[|g_{k,k'}|^2] \right ) \Id_T. \label{larsson-suka1}
\end{eqnarray}
It follows that 
\begin{eqnarray} 
h(y_k[1:T] |s_k[1:T])  & \leq & \EE \left [ \log (\pi e)^T \left | \left ( N_0 + \sum_{k'\neq k} \EE[|g_{k,k'}|^2] \right ) \Id_T + \sv_k \sv_k^\herm {\rm Var}(g_{k,k}) \right | \right ] \nonumber \\
& = & T \log (\pi e) + T \log \left ( N_0 + \sum_{k'\neq k} \EE[|g_{k,k'}|^2] \right ) \nonumber \\
& & + \EE \left [ \log \left ( 1 + \frac{{\rm Var}(g_{k,k}) \|\sv_k\|^2}{N_0 + \sum_{k'\neq k} \EE[|g_{k,k'}|^2]} \right ) \right ] \label{pippo3} \\
& \leq & T \log (\pi e) + T \log \left ( N_0 + \sum_{k'\neq k} \EE[|g_{k,k'}|^2] \right ) \nonumber \\
& & + \log \left ( 1 + \frac{T {\rm Var}(g_{k,k})}{N_0 + \sum_{k'\neq k} \EE[|g_{k,k'}|^2]} \right ) \label{larsson-suka2}
\end{eqnarray}
where in (\ref{pippo3}), after some simple manipulation, we used the fact that for, two $T$-length vectors $\av$ and $\bv$, 
$\log | \Id_T + \av \bv^\herm | = \log(1 + \bv^\herm \av)$, and in (\ref{larsson-suka2}
we applied Jensen's inequality to obtain simpler upper bound without expectations outside the log.

The second differential entropy in the RHS of (\ref{porcodio3}) can be lowerbounded by introducing conditioning (conditioning reduces the differential entropy 
\cite{cover2012elements}). We can write
\begin{eqnarray}
h(y_k[1:T] | g_{k,k}, s_k[1:T]) & \geq & h(y_k[1:T] | g_{k,k}, s_k[1:T], \{g_{k,k'} : k' \neq k\})  \nonumber \\
& = & T \log(\pi e) + T \EE\left [ \log \left ( N_0 + \sum_{k' \neq k} |g_{k,k'}|^2 \right ) \right ], \label{larsson-suka3}
\end{eqnarray}
where we used the fact that $\yv_k$ given $\sv_k$ and $\{g_{k,k'}\}$ has the same conditional differential entropy of the conditional Gaussian i.i.d. vector 
$\wv = \sum_{k' \neq k} \sv_{k'} g_{k,k'} + \zv_k$ given $\{g_{k,k'} : k' \neq k\}$. 
Using the upper bound (\ref{larsson-suka2}) and the lower bound (\ref{larsson-suka3}) in (\ref{porcodio3}), we obtain the upper bound
\begin{eqnarray}
I \left (g_{k,k}  ; y_k[1:T] | s_k[1:T]  \right ) & \leq & 
T \log \left ( N_0 + \sum_{k'\neq k} \EE[|g_{k,k'}|^2] \right ) - T \EE\left [ \log \left ( N_0 + \sum_{k' \neq k} |g_{k,k'}|^2 \right ) \right ] \nonumber \\
& & + \log \left ( 1 + \frac{T {\rm Var}(g_{k,k})}{N_0 + \sum_{k'\neq k} \EE[|g_{k,k'}|^2]} \right ).  \label{porcodio4}
\end{eqnarray}
Finally, using the lower bound (\ref{term13}) and the upper bound (\ref{porcodio4}) in (\ref{term03}), simplifying common terms and dividing by$T$, we 
obtain (\ref{LB3}). 
\hfill \QED

It is interesting to observe that bound (\ref{LB3}) has a much better behavior than bound (\ref{LB2}) for very large SNR, i.e., in the limit 
$N_0 \rightarrow 0$. In fact, for typical values of the coherence block size $T$ (see examples in Section \ref{results}), 
the second term in the RHS of (\ref{LB3}) is typically much smaller than the first term for all $N_0$ and, while 
the difference of the third and fourth terms is negative (by Jensen's inequality), it remains bounded as $N_0 \rightarrow 0$. 
This can be readily shown as follows: for $N_0 \rightarrow 0$, we can write the Jensen's penalty term
\begin{align} 
\log \left ( 1 + \frac{1}{N_0} \sum_{k'\neq k} \EE[|g_{k,k'}|^2] \right )  - \EE\left [ \log \left ( 1 + \frac{1}{N_0} \sum_{k'\neq k} |g_{k,k'}|^2 \right ) \right ]  &  \nonumber \\
= \log \left ( \sum_{k'\neq k} \EE[|g_{k,k'}|^2] \right )  - \EE\left [ \log \left ( \sum_{k'\neq k} |g_{k,k'}|^2 \right ) \right ]   + O(N_0) 
\end{align}
where the difference in the RHS is positive but independent of $N_0$. 
In particular, when these coefficients do not display large fluctuations (e.g., for the most useful fading statistics and channel estimation schemes the channel coefficients
have bounded moments of all orders), the Jensen's penalty is typically small. 
}

We conclude this section with an immediate extension of Lemmas \ref{LB1-lemma}, \ref{LB2-lemma}, and \ref{LB3-lemma}  to the case of 
receiver side information. Suppose that each user $k$ has a receiver side information $\Omega_k[t]$ such that 
\[ \{\Hm[t], \vv_1[t], \ldots, \vv_K[t], \Ec_1[t], \ldots, \Ec_K[t], \Omega_1[t], \ldots, \Omega_K[t]\} \]
is jointly stationary and ergodic and  and $\Omega_k[t]$ is independent of the transmitted codewords 
$\{ s_{k'}[tT; (t+1)T-1] : k' = 1, \ldots, K\}$.
This means that the receiver side information conveys to each receiver $k$ only information about the effective channel
coefficients, and not on the information messages of the users, i.e., it cannot be used to improve decoding through (partial) 
interference cancellation. Such an assumption is indeed realistic and relevant to the case of MU-MIMO DL beamforming, 
capturing the case where the side information is obtained through some
pilot scheme to learn the channel coefficients. Then, the achievable rate is given by 
\begin{equation} \label{sideinfo}
R_k = \frac{1}{T} I(s_k[1:T]; y_k[1:T] | \Omega_k), 
\end{equation}
where $\Omega_k$ has the same first-order marginal distribution of $\{\Omega_k[t]\}$. In this case, the bounds (\ref{LB1}), (\ref{LB2}), and (\ref{LB3})  
are modified as
\begin{equation}  \label{LB1-sideinfo}
R^{\rm lb1}_k = \EE \left [ \log \left ( 1 + \frac{\left | \EE\left [ g_{k,k} | \Omega_k \right ] \right |^2 }
{N_0 +  {\rm Var} \left ( g_{k,k} |\Omega_k \right )
+ \sum_{k' \neq k} \EE \left [ |g_{k,k'}|^2 |\Omega_k \right ]}  \right ) \right ], 
\end{equation}
\begin{equation}  \label{LB2-sideinfo}
R^{\rm lb2}_k = 
\EE \left [  \log \left ( 1 + \frac{|g_{k,k}|^2}{N_0 + \sum_{k' \neq k} |g_{k,k'}|^2}    \right ) \right ]  
- \frac{1}{T} \sum_{k'=1}^K \EE\left [ \log \left (1 + \frac{T}{N_0} {\rm Var}(g_{k,k'} | \Omega_k ) \right ) \right ],
\end{equation}
and
{\BLUE
\begin{align}  \label{LB3-sideinfo}
R^{\rm lb3}_k & = 
\EE \left [ \log \left ( 1 + \frac{|g_{k,k}|^2}{N_0 + \sum_{k' \neq k} \EE[ |g_{k,k'}|^2 | g_{k,k}, \Omega_k] }    \right ) \right ]  
- \frac{1}{T} \EE \left [ \log \left (1 + \frac{T {\rm Var}(g_{k,k} | \Omega_k)}{N_0 + \sum_{k' \neq k} \EE[|g_{k,k'}|^2 | \Omega_k]}  \right ) \right ] \nonumber \\
& + \EE\left [ \log \left ( 1 + \frac{1}{N_0} \sum_{k'\neq k} |g_{k,k'}|^2 \right ) \right ]  -  \EE \left [ \log \left ( 1 + \frac{1}{N_0} \sum_{k'\neq k} \EE[|g_{k,k'}|^2|\Omega_k] \right ) \right ] 
\end{align}
}
respectively, where for two random variables $X$ and $Y$ we define the conditional variance as~\footnote{Notice that 
the symbol ${\rm Var}(X|Y)$ is often used with an ambiguous meaning. In some textbooks it is defined as
${\rm Var}(X|Y) = \EE \left [ | X - \EE[X|Y]|^2 \right ]$, i.e., it is not a function of the conditioning variable $Y$, inconsistently with
the definition of conditional expectation $\EE[X|Y]$ which is indeed a function of the conditioning variable \cite{grimmett2001probability}.
To avoid misunderstanding, we gave an explicit definition.} 
\begin{equation} 
{\rm Var}(X|Y) = \EE \left [ | X - \EE[X|Y]|^2 | Y \right ] = \EE[|X|^2|Y] - |\EE[X|Y]|^2. 
\end{equation}
The proof of (\ref{LB1-sideinfo}) -- (\ref{LB3-sideinfo}) follows in the footsteps of the proofs of 
Lemmas \ref{LB1-lemma} -- \ref{LB3-lemma}, and it is omitted for the sake of brevity.

\section{Examples}  \label{results}

\begin{figure}[ht]
\centerline{\includegraphics[width=8cm]{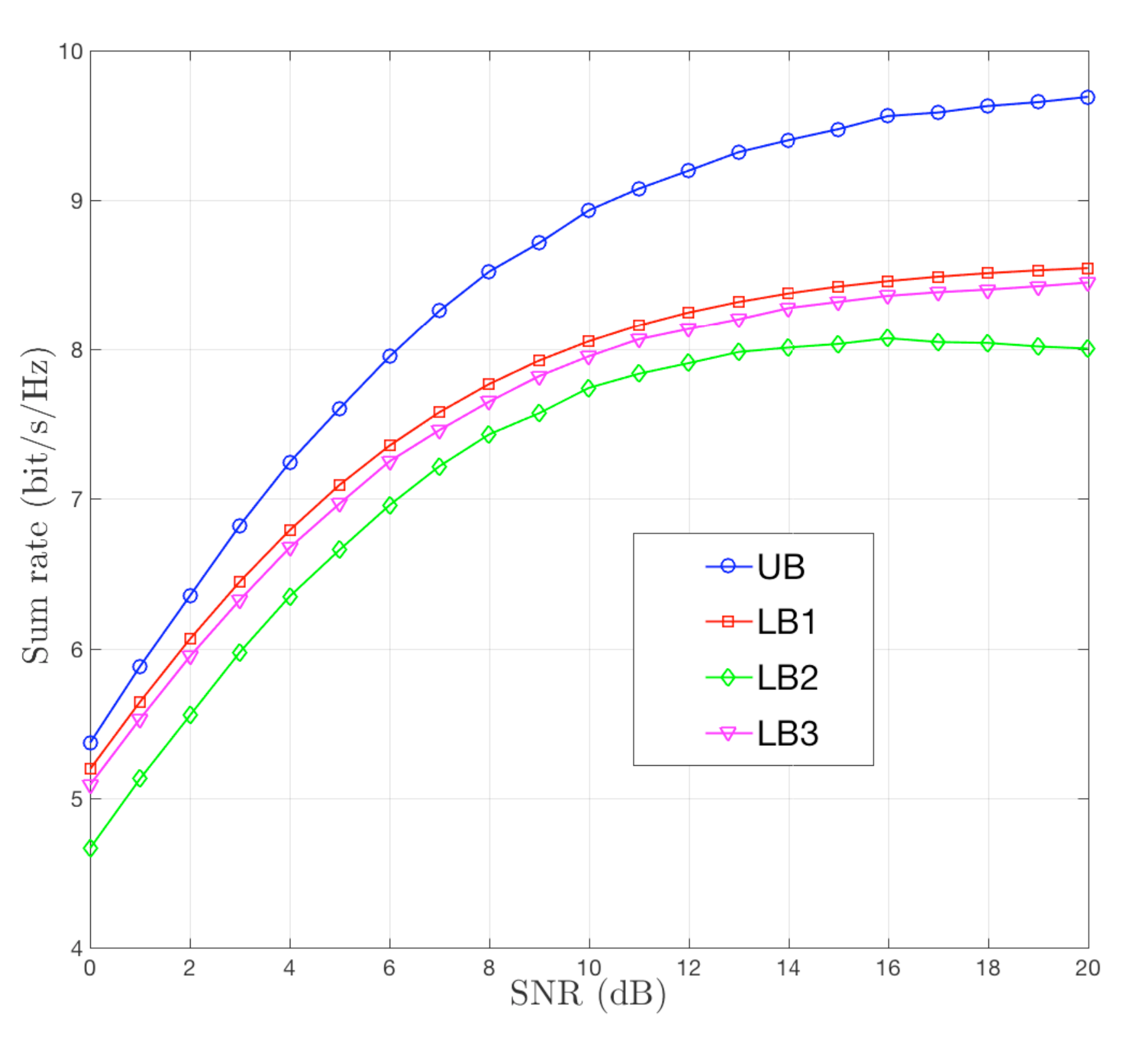} \hspace{1cm} \includegraphics[width=8cm]{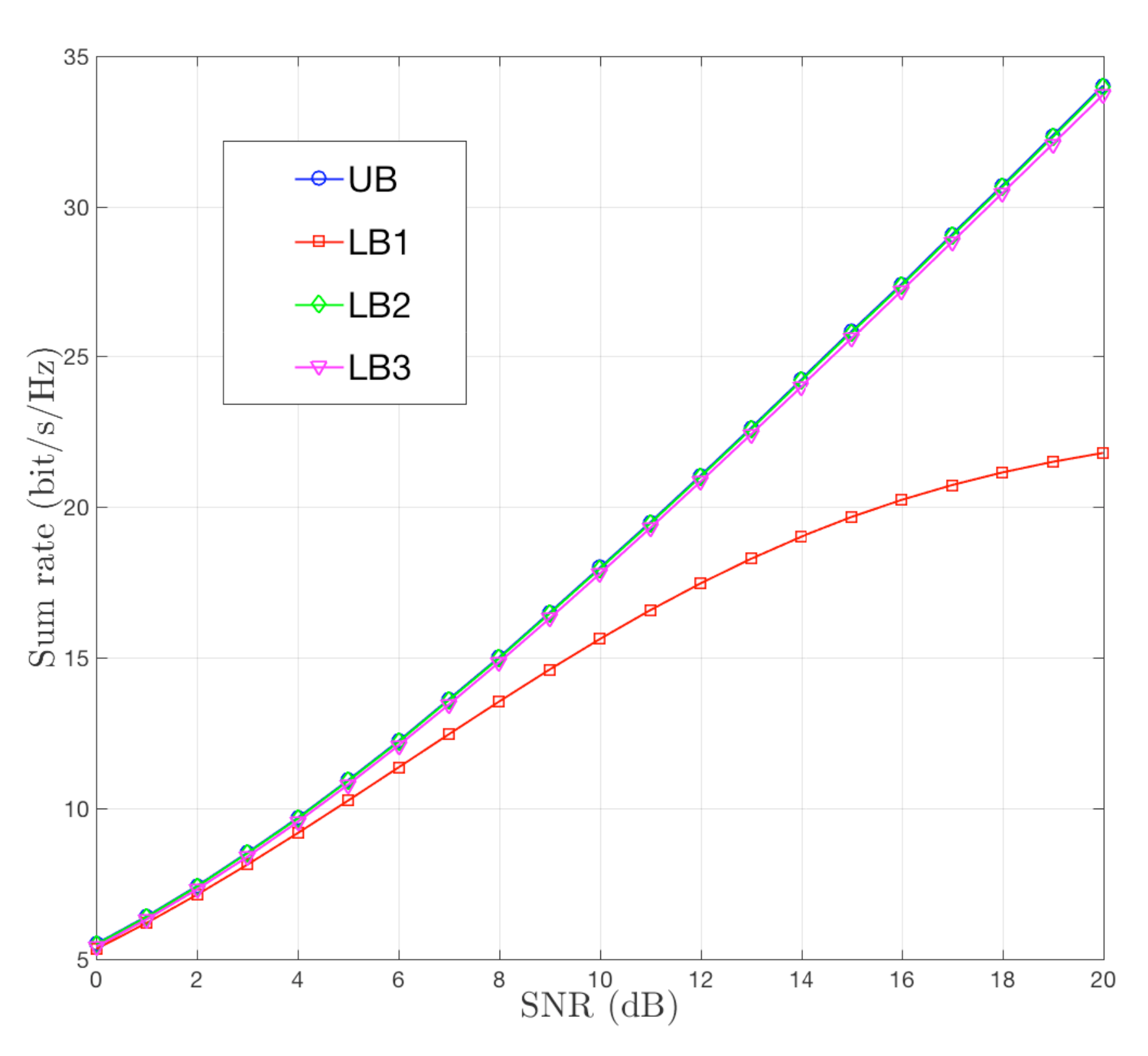}}
\caption{Ergodic sum rate for a system with $M = 10$ base station antennas, $K = 5$ users, and coherence time $T = 168$. 
The precoding vectors are calculated from the exact knowledge of the channel matrix. 
The chart on the left shows the performance of conjugate beamforming (ConjBF), and the plot on the right shows the performance of 
zero-forcing beamforming (ZFBF).}
\label{M10K5T100-analytic}
\end{figure}

{\BLUE
For the sake of simplicity, we consider the classical case where $\Hm[t]$ is Gaussian i.i.d. with elements 
$\sim \Cc\Nc(0,1)$.
First, we consider the case of perfect CSI at the base station, such that the DL beamforming vectors
can be computed from $\Hm[t]$.  Even in this case, each receiver $k$ does not know a priori the effective channel coefficients 
$\{g_{k,k'}\}$. Hence, we resort to the bounds in order to evaluate the achievable ergodic rate. 
With reference to the channel model (\ref{ziominchia}), 
with Conjugate Beamforming (ConjBF), the beamforming vectors are given by 
$\vv_{k'} = \hv_{k'}/\|\hv_{k'}\|$.  With equal energy per symbol per data stream,  the effective channel coefficients are given by 
\[ g_{k,k'} = \left \{ \begin{array}{ll}
\sqrt{ \frac{\Ec_{\rm tx}}{K}} \|\hv_k\|  & \mbox{for} \;\; k' = k \\
\sqrt{\frac{\Ec_{\rm tx}}{K}}  \hv_k^\herm \hv_{k'} /\|\hv_{k'}\| & \mbox{for} \;\; k' \neq k \end{array} \right . \]
Notice also that 
\begin{subequations}
\begin{eqnarray} 
\EE[ |g_{k,k'}|^2 | g_{k,k} ] & = & \frac{\Ec_{\rm tx}}{K} \EE \left [ \left . \frac{|\hv_k^\herm \hv_{k'}|^2}{\|\hv_{k'}\|^2} \right | \|\hv_k\| \right ] \nonumber \\
& = & \frac{\Ec_{\rm tx}}{K} \|\hv_k\|^2 \EE [  |\uv^\herm \vv|^2] \label{uv} \\
& = & \frac{\Ec_{\rm tx}}{K} \Xc_{2M}  \EE[ \beta_{1,M-1} ] =  \frac{\Ec_{\rm tx}}{MK} \Xc_{2M},
\end{eqnarray}
\end{subequations}
where $\uv,\vv$ denote two independent $M$-dimensional unit vectors, we used the fact that $|\uv^\herm \vv|^2$ is distributed as 
$\beta_{1,M-1}$, a beta-distributed random variable with
parameters $1$ and $M-1$, $\EE[\beta_{1,M-1}] = 1/M$, 
and $\Xc_{2t}$ denotes a central chi-squared random variable with $2t$ degrees of freedom. 
Therefore, the first term in (\ref{LB3}) can be written as
\begin{eqnarray} 
\EE \left [  \log \left ( 1 + \frac{|g_{k,k}|^2}{N_0 + \sum_{k' \neq k} \EE[ |g_{k,k'}|^2 | g_{k,k}] }    \right ) \right ]  & = & 
\EE \left [  \log \left ( 1 + \frac{\Xc_{2M}}{K N_0/\Ec_{\rm tx} +  \frac{K-1}{M} \Xc_{2M}}    \right ) \right ] \nonumber \\
& = & \EE \left [ \log \left ( 1 + \gamma_1 \Xc_{2M} \right ) - 
\log \left ( 1 + \gamma_2 \Xc_{2M} \right ) \right ] \nonumber \\
& = & \left ( \Ic_{M}(\gamma_1) - \Ic_{M}(\gamma_2) \right ) \log(e) 
\label{arwen}
\end{eqnarray}
where we used the fact that $\|\hv_k\|^2$ is central chi-squared with $2M$ degrees of freedom, 
where we define the integral
\begin{eqnarray} 
\Ic_n (\mu) & = & \EE[\log_e(1 + \mu \Xc_{2M})] \nonumber \\
& = & \Pi_M(-1/\mu) {\rm E_i}(1,1/p) + \sum_{m=1}^{M-1} \frac{1}{m} \Pi_m(1/\mu) \Pi_{M-m}(-1/\mu)
\end{eqnarray}
where
\[ \Pi_n(x) =  e^{-x} \sum_{i=0}^{n-1} \frac{x^i}{i!}, \;\;\;\; {\rm E_i}(n,x) = \int_1^\infty \frac{e^{-xt}}{t^n} dt   \]
where we defined the coefficients
\begin{eqnarray}
\gamma_1 & = & \frac{M+K-1}{MK} \frac{\Ec_{\rm tx}}{N_0} \nonumber \\
\gamma_2 & = & \frac{K-1}{MK} \frac{\Ec_{\rm tx}}{N_0},
\end{eqnarray}
and where we used the result in \cite[Eq. 7]{caire1999optimum}. 
While in general it is difficult to give a closed-form for the upper bound (\ref{UB}), which coincides with the first term in the lower bound (\ref{LB2}), 
we notice that it is quite easy to accurately evaluate this term by Monte Carlo simulation. Furthermore, as far as the lower bound (\ref{LB2}) is concerned, 
we can slightly relax it by replacing the first term by (\ref{arwen}), which provides indeed a lower bound as shown as a simple application of conditioning and Jensen's inequality 
as follows
\begin{eqnarray}
\EE \left [  \log \left ( 1 + \frac{|g_{k,k}|^2}{N_0 + \sum_{k' \neq k} |g_{k,k'}|^2 }    \right ) \right ]  & = & 
\EE \left [ \EE \left [  \left . \log \left ( 1 + \frac{|g_{k,k}|^2}{N_0 + \sum_{k' \neq k} |g_{k,k'}|^2 }    \right ) \right | \hv_k \right ] \right ]  \nonumber \\
& \geq &  \EE \left [  \log \left ( 1 + \frac{\Xc_{2M}}{KN_0/\Ec_{\rm tx} + (K-1) \Xc_{2M} \EE[ \beta_{1,M-1}] }    \right ) \right ] \nonumber \\
& & \label{arwen1}
\end{eqnarray}
where we recognize that (\ref{arwen1}) is given again by (\ref{arwen}).  

In order to compute the other terms in the bounds for the case of ConjBF we need the following immediate results
\begin{subequations}
\begin{eqnarray}
\EE[g_{k,k}] & = & \sqrt{\frac{\Ec_{\rm tx}}{K}} \frac{\Gamma(M + 1/2)}{\Gamma(M)}, \label{millo}\\
\EE[|g_{k,k}|^2] & = & \frac{\Ec_{\rm tx}}{K} M \label{millo1} \\
\EE[g_{k,k'}] & = & 0, \;\;\; \forall \; k' \neq k \\
\EE[|g_{k,k'}|^2] & = & \frac{\Ec_{\rm tx}}{K}.
\end{eqnarray}
\end{subequations}
Notice that the third term in (\ref{LB3}) is not amenable to a closed-form expression and must also be computed by Monte Carlo simulation. 
}
In the case of zero-forcing beamforming (ZFBF), 
the base station calculates the (unit-norm) precoding vectors as the normalized columns of the Moore-Penrose channel matrix pseudo-inverse
$\Hm (\Hm^\herm \Hm)^{-1}$. In this case, it is a simple matter to show that
\[ g_{k,k'} = \left \{ \begin{array}{ll}
 \sqrt{\frac{\Ec_{\rm tx}}{K}} \|\tilde{\hv}_k\|  & \mbox{for} \;\; k' = k \\
0 & \mbox{for} \;\; k' \neq k \end{array} \right . \]
where $\widetilde{\hv}_k$ is a vector with $M - K + 1$ independent components $\sim \Cc\Nc(0,1)$. 
It follows that  $\|\widetilde{\hv}_k \|^2$ is distributed as $\Xc_{2(M-K+1)}$, and (\ref{UB}), as well as the first term in (\ref{LB2}) and in (\ref{LB3}) are given in closed form as
{\BLUE
\begin{equation} 
\EE\left [ \log \left (1 + \gamma_3 \Xc_{2(M-K+1)} \right ) \right ] =  \Ic_{M-K+1}(\gamma_3) \log(e) 
\end{equation}
with $\gamma_3 = \Ec_{\rm tx}/(K N_0)$. 
In order to evaluate the lower bounds we need also the mean and second moment of the useful signal coefficient, given by 
\begin{subequations}
\begin{eqnarray}
\EE[g_{k,k}] & = & \sqrt{\frac{\Ec_{\rm tx}}{K}} \frac{\Gamma(M-K+3/2)}{\Gamma(M-K+1)} \\
\EE[|g_{k,k}|^2] & = & \frac{\Ec_{\rm tx}}{K} (M - K + 1) 
\end{eqnarray}
\end{subequations}

Fig.~\ref{M10K5T100-analytic} shows the sum ergodic rate bounds 
as a function of $\SNR \eqdef \Ec_{\rm tx}/N_0$, for a system with $M = 10$ antennas, $K = 5$ users, 
ideal knowledge of the matrix channel matrix for the calculation of the precoding vectors, i.i.d. channel coefficients $\sim \Cc\Nc(0,1)$
and equal power allocation, corresponding to the above expressions. 
We used channel coherence block $T = 14 \times 12 = 168$ signal dimensions, motivated by the size of an LTE resource block, 
that spans $14$ OFDM symbols in time and $12$ adjacent subcarriers in frequency. 
We notice that in the case of ConjBF (see also the other figures in this section) 
the best lower bound is always provided by LB1 (Lemma \ref{LB1-lemma}). 
In contrast, LB1 it is significantly outperformed by LB2 and LB3 (resp., Lemma \ref{LB2-lemma} and \ref{LB3-lemma})
for the case of  ZFBF,  where LB1 displays a ``self-interference limited'' behavior.  
We explain this fact by noticing that in the case of ConjBF and a small number of antennas ($M = 10$ in this case)
the system is heavily interference limited and the self-interference term Var$(g_{k,k})$ in the denominator of (\ref{LB1}) 
is negligible with respect to the multiuser interference term $\sum_{k' \neq k} \EE[|g_{k,k'}|^2]$. 
This is no longer true for the case of ZFBF, where interference is removed by zero-forcing beamforming. 

It was noticed in Section \ref{rate-bounds} that LB2 may become useless (indeed, negative) for very large SNR. 
This case only occurs when the first term in (\ref{LB2}) is interference-limited. 
In the case of perfect channel matrix knowledge at the transmitter, 
this term is interference limited for the case of ConjBF, while it is not in the case of ZFBF, since $g_{k,k'} = 0$ for all $k' \neq k$. 
In order to evaluate the rate penalty caused by the second term in 
(\ref{LB2}), assume that the terms Var$(g_{k,k'})$ are all equal to $\Ec_{\rm tx}/N_0$.\footnote{It can be checked from (\ref{millo}) and (\ref{millo1}) 
that variance of the coefficient $g_{k,k}$ is significantly smaller, but we use this approximate argument in order to provide a simple calculation.}
Then, the second term in (\ref{LB2}) is given by 
$\frac{K}{T} \log \left ( 1 + \frac{T}{K} \frac{\Ec_{\rm tx}}{N_0} \right )$. 
 For example, for $T = 168$ and $K = 5$ as in Fig.~\ref{M10K5T100-analytic}, 
at $\SNR = 10$ dB the rate penalty is $\approx 0.25$ bits. Notice that UB at 10 dB yields a sum rate of $\approx 9$ bits, 
i.e., $1.8$ bit per user. Hence, at 10 dB the penalty with respect to the upper bound is already $\geq14\%$ of the optimal achievable rate.
This shows that LB2 does not give meaningful results for heavily interference-limited systems. 
Nevertheless, letting $T \rightarrow \infty$ in LB2 we have immediately that the upper bound UB is achievable in the limit of large block length. 
Although the achievability of the ``genie-aided'' (perfect CSI knowledge) receiver in the non-coherent block-fading channel
in the limit of large $T$ is well-known (see \cite{BPS}), we found it nice that it follows so easily as an application of LB2.  
} 

\begin{figure}[ht]
\centerline{\includegraphics[width=8cm]{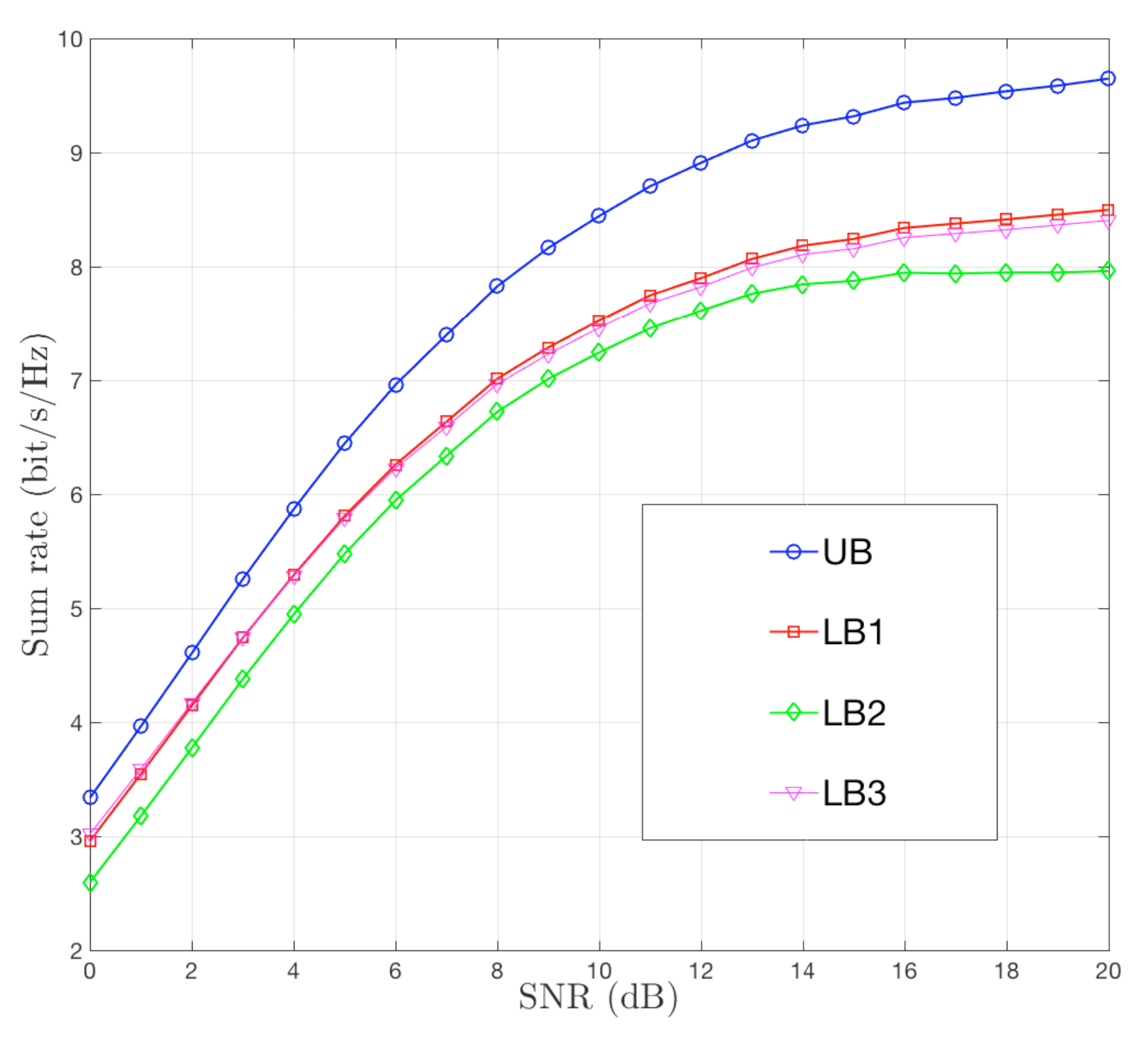} \hspace{1cm} \includegraphics[width=8cm]{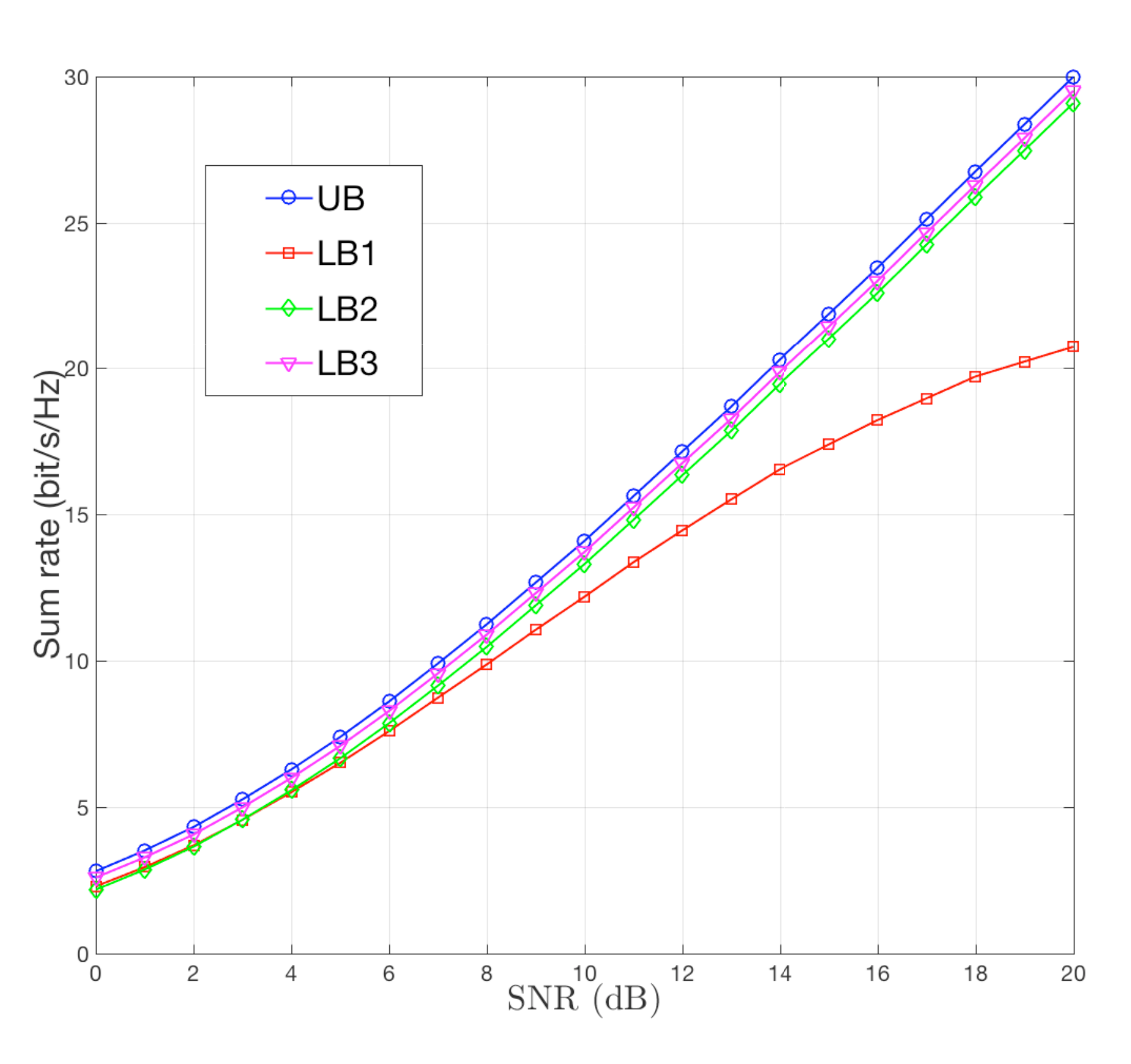}}
\caption{Ergodic sum rate for a system with $M = 10$ base station antennas, $K = 5$ users, and coherence time $T = 168$. 
The precoding vectors are calculated from a noisy observation of the channel matrix obtained through TDD reciprocity.
The chart on the left shows the performance of conjugate beamforming (ConjBF), and the plot on the right shows the performance of 
zero-forcing beamforming (ZFBF).}
\label{M10K5T100-noisyCSIT}
\end{figure}

Fig.~\ref{M10K5T100-noisyCSIT} shows analogous results for the case where the precoding vectors are calculated from a noisy observation of
the channel matrix, as obtained from TDD reciprocity via orthogonal uplink pilot symbols. The SNR for the pilot observation is also equal to 
$\Ec_{\rm tx} / N_0$. This assumption justified as follows: we assume that each user transmit in the uplink with the same 
energy per symbol $\Ec_{\rm tx}/K$ per user used in the DL by the base station. However, since the uplink pilots consist of $K$ channel uses, 
in order to make $K$ mutually orthogonal pilot sequences, the total pilot energy is $K$ times the energy per symbol, therefore the 
signal to noise ratio for the uplink channel estimation is given by $\SNR = \Ec_{\rm tx} / N_0$.

\begin{figure}[ht]
\centerline{\includegraphics[width=8cm]{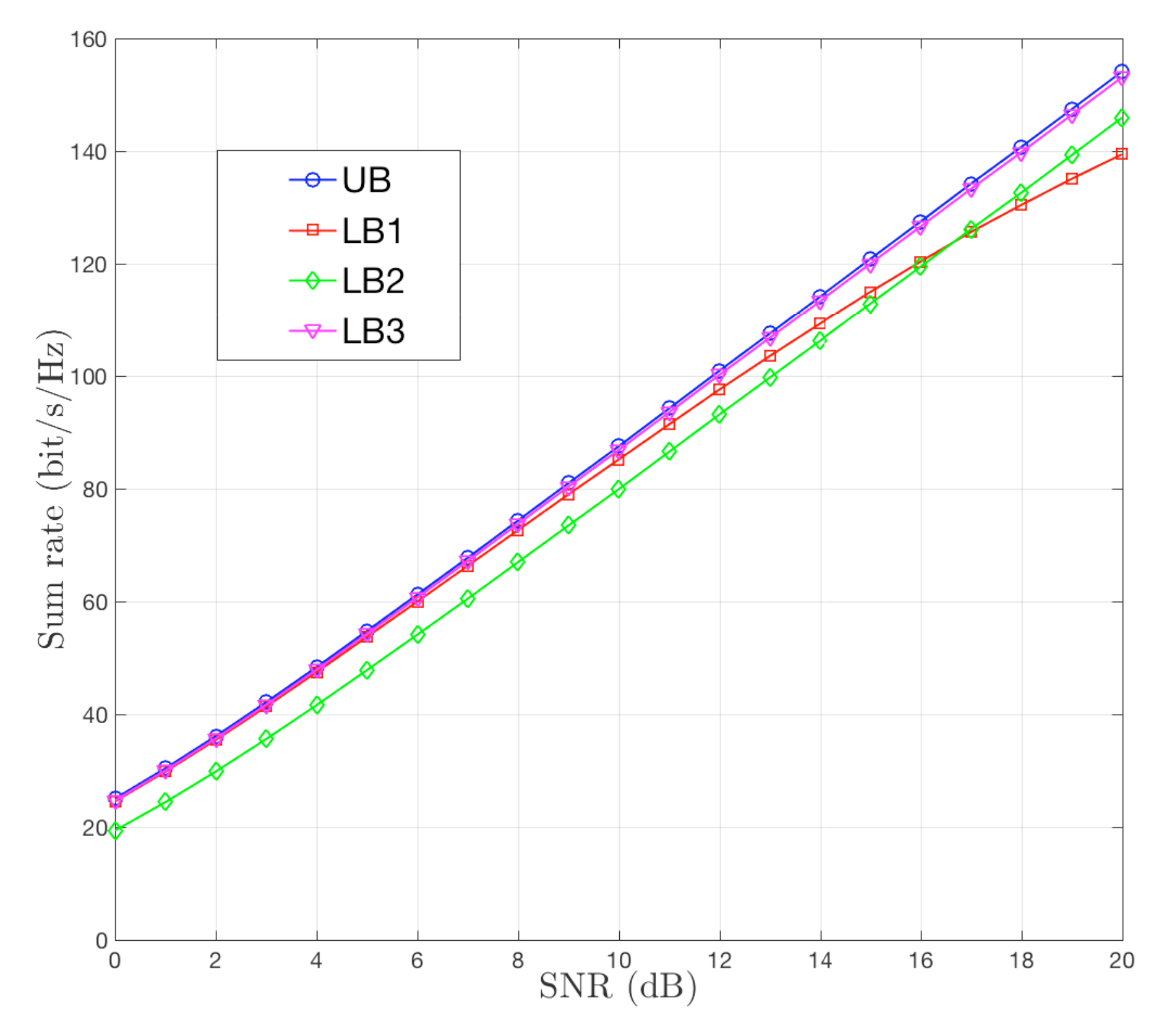} \hspace{1cm} \includegraphics[width=8cm]{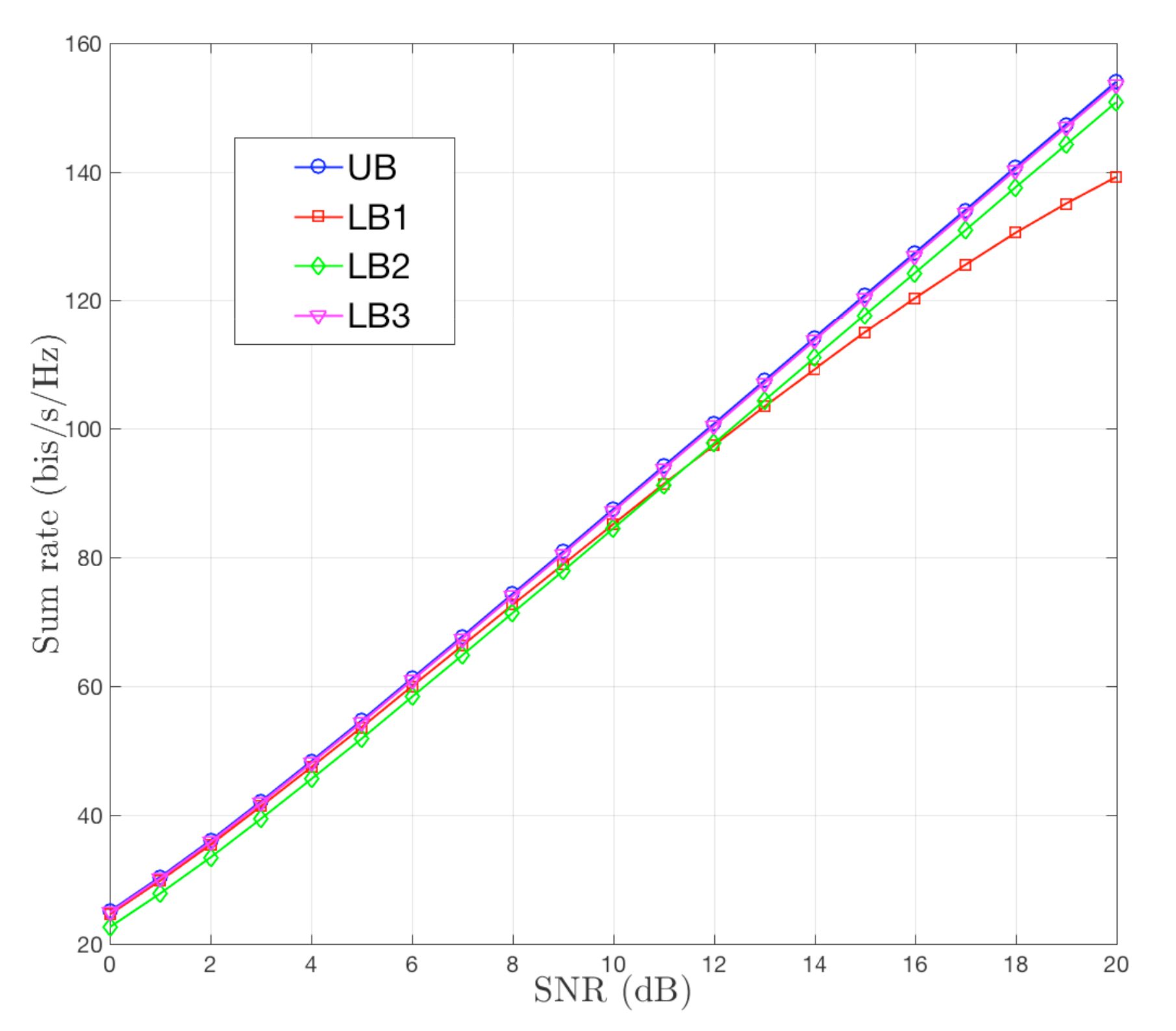}}
\caption{Ergodic sum rate for a system with $M = 100$ base station antennas, $K = 20$ users, ZFBF,  and coherence time $T = 168$ (left) 
and $T = 4 \times 168 = 672$ (right).  
The precoding vectors are calculated from a noisy observation of the channel matrix obtained through TDD reciprocity.}
\label{M100K20T100-noisyCSIT}
\end{figure}

{\BLUE 
The channel matrix observation from the transmission of $K$ orthogonal uplink pilots is given by 
\begin{equation} 
\Ym^{\rm ul} = \Hm + \sqrt{\frac{N_0}{\Ec_{\rm tx}}} \Wm,
\end{equation}
where $\Hm = [\hv_1, \ldots, \hv_K]$ is the channel matrix and $\Wm$ is a matrix with i.i.f. components $\sim \Cc\Nc(0,1)$. 
The base station finds an estimate $\widehat{\Hm}$ of $\Hm$ using linear MMSE estimation, given by 
\begin{equation}
\widehat{\Hm} = \frac{\Ec_{\rm tx}}{\Ec_{\rm tx}+ N_0} \Ym^{\rm ul}.
\end{equation}
The base station computes the ConjBF and the ZFBF beamforming vectors from the channel matrix estimate $\widehat{\Hm}$. 
In particular,  in the case of ConjBF we have $\vv_{k'} = \widehat{\hv}_{k'} / \|\widehat{\hv}_{k'}\|$ where $\widehat{\hv}_{1}, \ldots, \widehat{\hv}_K$ 
are the columns of $\widehat{\Hm}$, while in the case of ZFBF we have that the precoding vectors are  given by the
normalized columns of  $\widehat{\Hm} (\widehat{\Hm}^\herm \widehat{\Hm})^{-1}$. 


In the case of imperfect CSI, obtaining closed form expressions for the terms in the bounds seems to be 
difficult if not impossible, with the exception of LB1, that depend only on first and 
second moments of the effective channel coefficients, that can be still easily calculated. This represents indeed a non-trivial advantage 
of LB1, that fully justifies its wide use in the massive MIMO literature. 
Also in the case of non-ideal CSI, we notice from Fig.~\ref{M10K5T100-noisyCSIT} that LB1 performs best in the ConjBF case, while
it is severely interference limited in the ZFBF case.

Fig.~\ref{M100K20T100-noisyCSIT} shows  the performance of ZFBF for system parameters more representative of massive MIMO, namely,  
$M = 100$, $K = 20$, $T = 168$ (left) and $T = 4 \times 168 = 672$ (right). The ZFBF precoding vectors are calculated from a noisy observation of
the channel matrix, as described before. 
We notice that in this case LB1  yields a significantly better behavior for a larger range of SNR, however, for high SNR the self-interference term become relevant
and the curve flattens out and separates from the UB. LB2 follows UB but $T = 168$ is too small with respect to $K = 20$ and therefore the gap
of LB2 with respect to UB is significant. Of course, this gap reduces by increasing $T$, as shown in the $T = 672$ example. 
Overall, LB3 yields the best bound for ZFBF, but its numerical evaluation is slightly more difficult.}

\section{Concluding remarks}

{\BLUE
In this paper we have provided two new lower bounds to the ergodic rate of a channel with 
noise and interference, where the channel coefficients change over time in a block-wise jointly ergodic and stationary fashion, 
while they stay constant over blocks of $T$ signal dimensions (coherence block), and where the receiver treats interference as noise. 
For the sake of direct comparison and for completeness, we also included
an upper bound (in the max-min sense) and a well-known lower bound that is widely used in the massive MIMO literature. 
These bounds find their main application in providing tractable expressions (closed-form or easily evaluated by Monte Carlo simulation)
for the ergodic rate of the users in a MU-MIMO system with any form of linear precoding. In particular, we believe that they can be useful to
analyze the performance of massive MIMO in the regime where channel hardening is not very strong, such that there is significant random fluctuation
of the effective channel coefficients after DL beamforming, but the channel coherence block length $T$ is large with respect to the number of served 
users $K$. Recent examples of such situations have been shown in the case of cell-free 
massive MIMO \cite{chen2017channel,ngo2015cell},  and in the case of highly correlated channel vectors \cite{mahdi-FDD,adhikary2013joint,adhikary2014joint,nam2017role,rao2014distributed,fang2017low}.

It is apparent from (\ref{LB3}) that the ``hardening'' of the useful signal coefficient $g_{k,k}$ is not needed in order to obtain a large
ergodic rate, as long as the coherence block $T$ is not too small. From an operational viewpoint, this indicates that no explicit DL beamformed 
pilot symbols are needed in order to achieve good rates on the  ``block-wise non-coherent'' channel given by (\ref{ziominchia}).
This observation may provide a motivation to devote some renewed interest in coding and modulation schemes for the non-coherent 
block-fading channel (e.g., see \cite{marzetta1999capacity,divsalar1990multiple}), or possible alternative consisting of ``plug-in'' approaches 
that estimate the useful signal coefficient in a blind way,  as investigated in \cite{Larsson-blind}. 
}

\bibliographystyle{IEEEtran}
\bibliography{IEEEabrv,refs}

\end{document}